\documentclass[aps,prb,groupedaddress,showpacs,twocolumn]{revtex4}%
\usepackage{bm}
\usepackage{psfig}%
\usepackage{amsmath}%
\usepackage{amsfonts}%
\usepackage{amssymb}%
\usepackage{graphicx}
\begin{document}
\title{Energy controlled insertion of polar molecules in dense fluids}
\author{Gianni De Fabritiis}
\email[]{g.defabritiis@ucl.ac.uk}
\author{Rafael Delgado-Buscalioni}
\email[]{r.delgado-buscalioni@ucl.ac.uk}
\author{Peter V. Coveney}
\email[]{p.v.coveney@ucl.ac.uk}
\affiliation{Centre for Computational Science, Department of Chemistry, University College
London, 20 Gordon Street, WC1H 0AJ London, U.K.}
\date{\today}

\begin{abstract}
We present a method to search low energy configurations of polar molecules in
the complex potential energy surfaces associated with dense fluids.  The
search is done in the configurational space of the translational and
rotational degrees of freedom of the molecule, combining steepest-descent and
Newton-Raphson steps which embed information on the average sizes of the
potential energy wells obtained from prior inspection of the liquid structure.
We perform a molecular dynamics simulation of a liquid water shell which
demonstrates that the method enables fast and energy-controlled water molecule
insertion in aqueous environments.  The algorithm finds low energy
 configurations of incoming water molecules around
three orders of magnitude faster than direct random insertion. 
 This method is an
important step towards dynamic simulations of open systems and it may also
prove useful for energy-biased ensemble average calculations of the chemical
potential.
\end{abstract}
\maketitle

Many processes of physical, chemical and biological interest
 involve open systems which exchange matter with their surroundings.
Molecular dynamics (MD) and Monte Carlo (MC) simulations of these
systems  often require a method for molecule insertion and, therefore, 
a method for  searching
 configurations with prescribed (low) potential energy. Indeed, a randomly
 placed molecule is likely to overlap with pre-existing atoms, releasing into
 the system a very high amount of energy.

The most natural setting for these systems is  the grand
canonical (GC) ensemble.
 Several methods for GC simulations  
 require the location of energy cavities for insertion
 (such as cavity-biased methods for GCMC \cite{allen,adams75,Mezei87})
 or careful control of the solvent insertion energy in the case of GCMD \cite{Pet92,Pet-bov}.
 Mass, momentum and energy transfer are also a key feature of a class of hybrid methods
for non-equilibrium simulations which couple an open MD region with an
interfacing continuum-fluid-dynamics domain
\cite{BusH1,Flek00}. Open boundaries in such hybrid
schemes can avoid finite size effects in small MD simulation boxes
\cite{SandraBus}, thereby saving on computational time. 
These sort of open boundaries 
 could also be used to improve 
the closed ``water shells'' widely used to hydrate restricted subdomains \cite{SCAAS}
in many MD simulations of biological systems.

 Water insertion is also particularly
important in protein simulations. For instance, it is possible to study protein
unfolding via gradual water insertion in the protein's cavities
\cite{Goodfellow1,Goodfellow2}. On the other hand, water molecules buried in
protein cavities at very low energies are essential for protein
structure and function \cite{Hermans96,Ivo98,AQP-03}. Indeed,  some tools
for MD simulations (such as {\sc dowser} \cite{Hermans96}) are
specialised for water insertion in hydrophilic cavities, 
leaving empty however the larger hydrophobic cavities which frequently contain stable
yet disordered water molecules relevant to  protein
function\cite{Ivo98,cai03biochemistry}.

Several methods for the calculation of ensemble averages require sampling 
the potential energy released to the system upon insertion of a test molecule \cite{allen,bennett76,shing82,lu03}.
Examples include calculation of the chemical potential, hydration energies and pair distribution functions \cite{guillot91}. 
The applicability of these methods can be  expanded to dense fluids
using techniques  that bias the sampling towards low energy
configurations.  Some of these techniques, 
 such as cavity-biased \cite{jed00,poh96} or excluded volume map
\cite{dei89} sampling, are however hampered by the considerable amount of time needed to
find ``cavities'' where the test molecule 
could be inserted without overlapping with others. 
In fact, these cavities are just proxies to search low energy configurations 
 which could better be identified by an energy controlled  insertion method.

The algorithms for water insertion 
proposed in the literature usually involve rather lengthy steps
which  comprise three separate parts: location of a suitable ``cavity'', normally using an
expensive grid search with $O(10^6)$ different cells \cite{Mezei87,poh96,Pet92}; random insertion in the cavity, followed
by a large number of energy minimisation steps (either of the inserted
molecule \cite{Pet92,Hermans96} or of the entire system
\cite{Goodfellow1}) and, finally, thermostatting the whole system
over a  one to ten picoseconds period to extract the extra energy released
upon insertion.  
In this article, 
we present a method to locate low energy configurations of dense
liquids that allows  insertion of solvent molecules {\em on-the-fly}:
avoiding  expensive grid search, non-local energy minimisation and 
thermostatting steps.

On the potential energy surface, low energies are located inside energy wells
whose local minima span a relatively large range of energy values.  The main
idea of the present method is to reconstruct the energy landscape with a
limited number of probes by constraining the search to be {\it inside} the
energy wells.  In fact, any excursion outside the explored well 
implies the loss  of all the information accumulated on
the current well which is effectively
equivalent to a random restart.  Efficiency is obtained by minimising both the number of
probes needed to determine if the target energy is found within the well and
the number of explored wells per successful insertion.  The present
minimisation algorithm generalises non-trivially to multiple degrees of
freedom the \textsc{usher} algorithm for insertion of Lennard-Jones atoms
\cite{usher}.  It shares with some other global minimisation methods the
recipe of applying in turns random moves and local energy minimisation
\cite{levitt75,li87,saunders87}. However, it is distinguished from these
others in the way the minimisation is performed via a combined
steepest-descent and Newton-Raphson iterator which is tailored adaptively to
the structure of the potential energy landscape being searched.


The method uses local information on the gradient and 
the average  
size  of the potential wells, which are dependent on the
molecule's location and the  thermodynamic state respectively. 
The input parameters specify the maximum 
distance $\Delta R$ and rotation angle $\Delta \Theta$ that the incoming molecule can jump without exiting the 
current well together with   
a measure of the roughness of the potential energy surface  $\Delta E_R$.
The insertion algorithm  starts by selecting a random location for the centre of mass of the molecule and placing the
atoms at the equilibrium bond and angle positions in a random orientation.
The non-bonded potential energy of an incoming  molecule  is given by 
\begin{equation}
U = \frac{1}{2}\sum_{i \neq j} V_{LJ}( r_{ij}) 
   + \frac{1}{2} \sum_{i \neq j} V_C( r_{ij}),  
\label{energy}
\end{equation}
where $V_{LJ}$ and $V_C$ are the Lennard-Jones and Coulomb pair potentials respectively \cite{allen} and 
the index i runs over the atoms of the molecule and j over all other atoms, which remain fixed while inserting. 
The  energy   $E = 2 U$ released to the system upon 
insertion is computed and compared with the target energy $E_{T}$.
 The insertion succeeds once  the energy difference $\Delta E=E-E_{T}$
is less than a certain prescribed tolerance set here at $10^{-3}$ Kcal/mol.

It is likely that for the random starting configuration  $\Delta E$ will be 
a large positive value because there is a high chance that the inserted 
molecule will overlap with others. 
Then, the force ${\bf F}=\sum_i {\bf F}_i$ applied to the centre of mass ${\bf r}_{cm}$ and torque 
${\bf \tau} = \sum_i {\bf r}_{cm,i} \times {\bf  F}_{i}$ are used to compute the next 
displacement and rotation. Here, the index $i$ runs over the atoms of the inserted molecule and 
${\bf r}_{cm,i} = {\bf r}_i - {\bf r}_{cm}$. The molecule is translated by 
$\delta r = \min(\Delta E/F,\Delta R)$ where $F$ is the magnitude of the force 
on the centre of mass and $\Delta R$ is the maximum displacement. 
With the reference system  fixed to the  molecule, 
 we then compute the rotation angle around the torque axis 
$\delta \theta = \min(\Delta E / \tau,\Delta \Theta)$ and rotate the molecule around the centre of mass.
The resulting update rule is finally given by 
\begin{eqnarray}
{\bf r}^{n+1}_{cm} &=& {\bf r}^{n}_{cm}+ \frac{ {\bf F^{n} }}{F^{n}} \delta r,\nonumber\\
{\bf r}^{n+1}_{cm,i}  &=& {\cal R}_{{\bf \tau}^n} {\bf r}^{n}_{cm,i}, \label{updaterule}
\end{eqnarray}
where  ${\cal R}$ is the rotation matrix around the axis of torque of angle $\delta \theta$.
This is equivalent to
a first order steepest descent procedure for large energy differences and a
second order Newton method for energy close to the 
target energy \cite{usher}.
 The angular minimisation is stopped when 
the angle $\delta \theta$ is less than $1^{\circ}$  to avoid oscillations due to  the coupling of 
rotational and translational degrees of freedom.
If during the iterations
$\Delta E$ increases by more than $\Delta E_{R} $  then the current attempt is abandoned and a new random
configuration is generated. This provides a threshold to control the amount of time  spent searching in the well and 
the number of wells explored.

The insertion algorithm in Eq. (\ref{updaterule}) does not require a baroque implementation and indeed 
can be easily included in any molecular dynamics program. The code used here
is based on the serial version of a well established parallel
molecular dynamics code NAMD with the Charmm27 force field \cite{NAMD}, but it has been designed to interface 
easily with any other  serial or parallel MD code.
The search algorithm applies in general to  small polar molecules but given its importance we focus on  controlled insertion of 
water molecules in aqueous environments.
We use the TIP3P model for water, widely utilised in
biological simulations \cite{jorgensen83}. 
This water model is based on  three
interaction sites, bonds (O-H) and angle (H-O-H) being constrained rigidly or, in its flexible version
(used here), by a harmonic potential
with  equilibrium configurations of  0.96 {\AA} and $104.52^{\circ}$ respectively. 

\begin{figure}[tb]
\begin{center}
\centerline{\psfig{figure=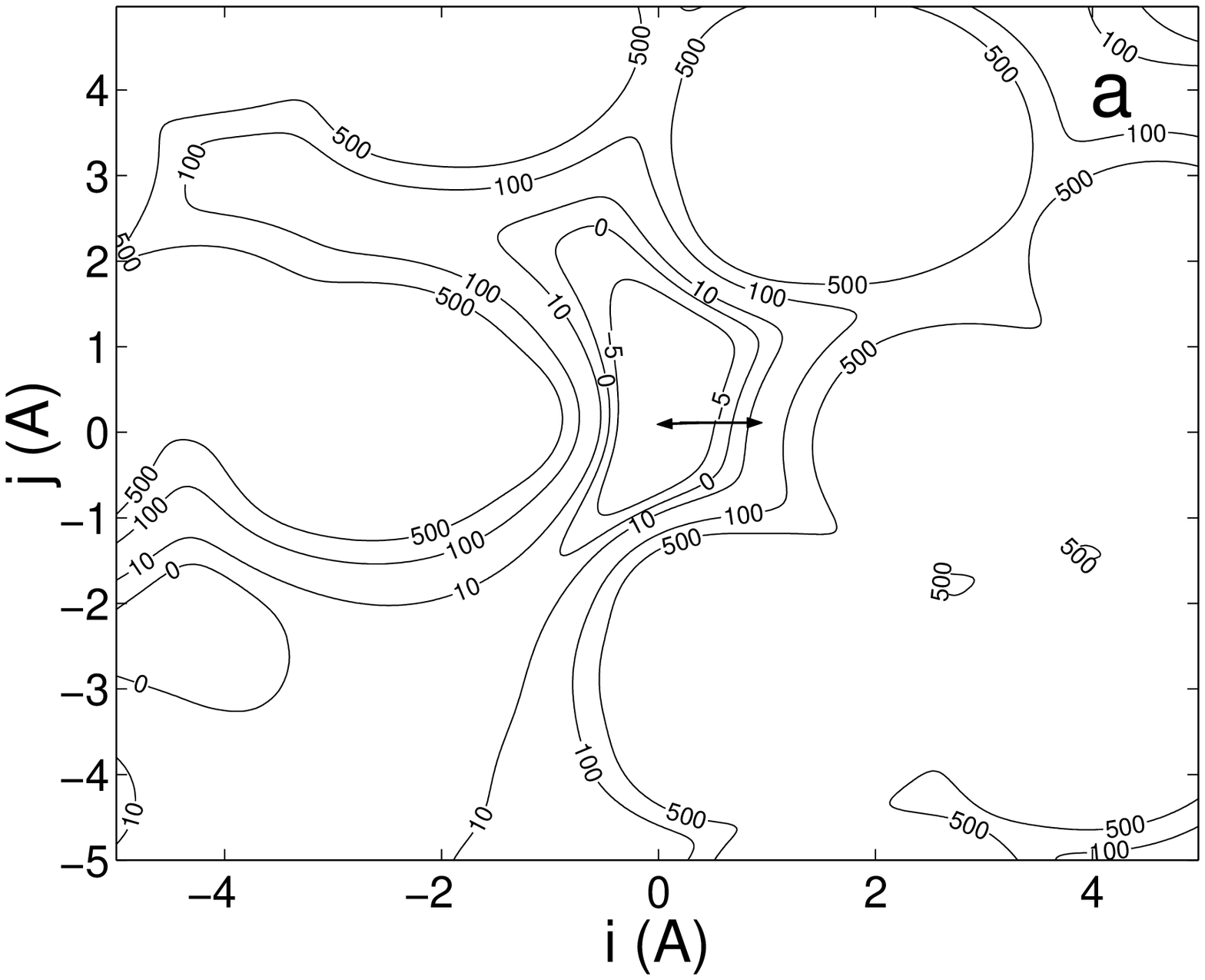,width=4.25cm}
\psfig{figure=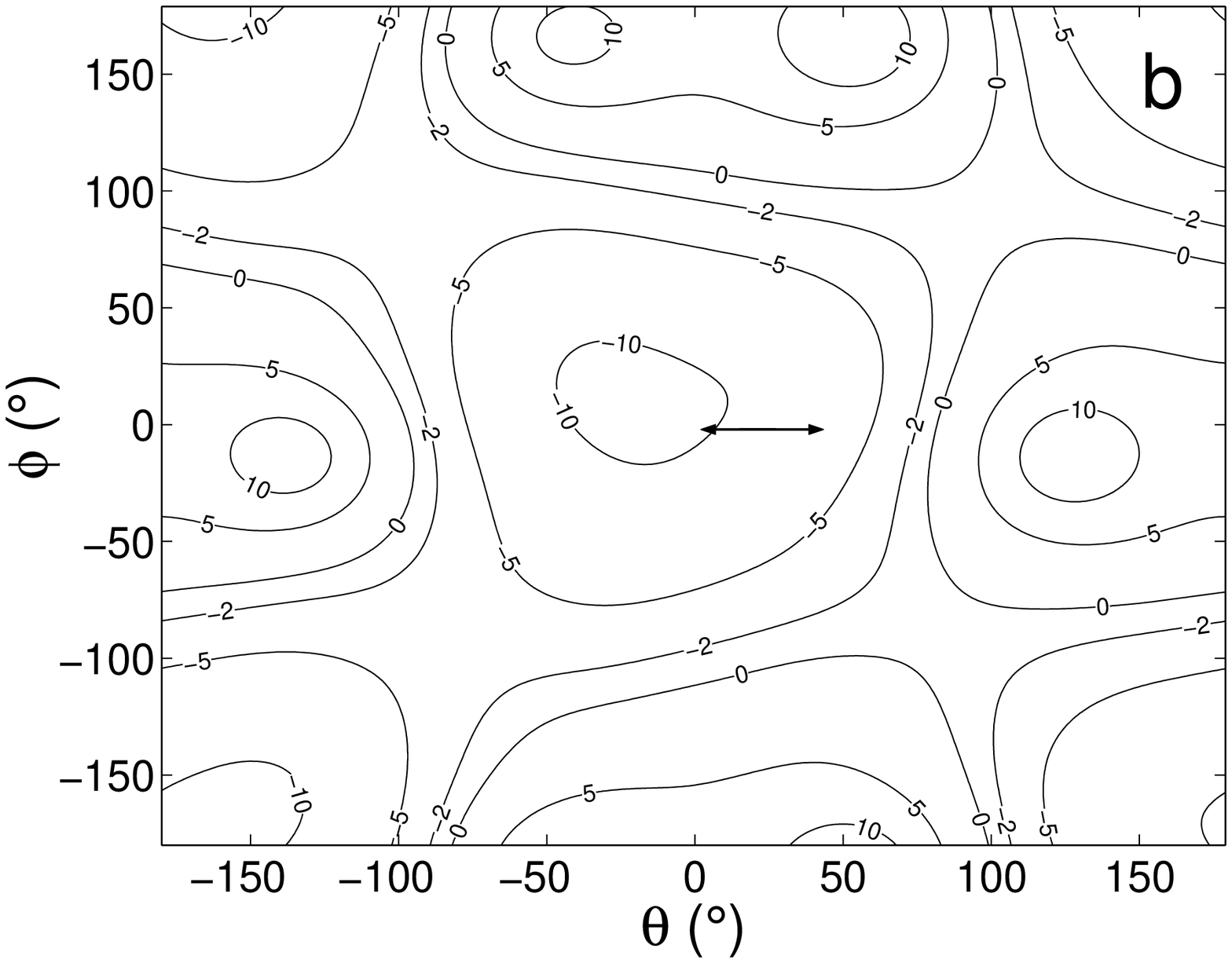,width=4.25cm}}
\end{center}
\caption{ Contour plot of the potential energy landscape in Kcal/mol: (a) for  translation  relative to the axes {\bf i}
and {\bf j} fixed to the water molecule; (b)  for a rotation $\theta$ about the axis {\bf j}
and $\phi$ about the axis {\bf k} for an equilibrated periodic liquid water system at 300K and 
density 0.96 $g/cm^{3}$. 
 The maximum translational   displacement $\Delta R = 1.0$ {\AA} and
   maximum rotational  angles  $\Delta \Theta = \Delta \Phi = 45^{\circ}$  are indicated by  double-headed arrows.
For visual convenience angles  smaller than $-90^{\circ}$ and larger than $90^{\circ}$ in $\theta$ are plotted although being redundant.}
\label{energy_ij}
\end{figure}

As stated, the restriction on the maximum displacement and rotation has the effect of limiting
the search to the current potential well.
For water, the maximum displacement  can be extracted from the oxygen-hydrogen pair distribution function $g_{OH}$ \cite{jorgensen83}. 
We found that an optimum value for the maximum displacement $\Delta R= 1$ {\AA} is  half of the first peak in $g_{OH}$ which  
is around 2 {\AA}. 
Exploring the potential energy landscape provides another simple way of obtaining the input parameters.
In Fig. \ref{energy_ij}a, we show a cross-section of the potential energy surface for a displacement of up to 5 {\AA}
around an equilibrated water molecule in the direction of the axes ${\bf i}$ and ${\bf j}$.
The unit vectors ${\bf i,j,k}$ form a reference system fixed rigidly to the water molecule 
with the axis ${\bf i}$ being in the direction of  the dipole.
As shown in Fig. \ref{energy_ij}a the optimum value of   $\Delta R$ is approximately   
 the radius of the  potential energy well, corroborating information furnished 
 from the pair distribution function.
It is more difficult to obtain structural information for the angular degrees of freedom.
However, a simple inspection of Fig. \ref{energy_ij}b provides a gross estimate of 
potential energy wells in the rotational degrees of freedom as being between $90-100^{\circ}$  wide; 
therefore the maximum rotation can be fixed at $\Delta \Theta = 45^{\circ}$.
The value of $\Delta E_{R}$, which sets the maximum uphill energy jump allowed in one move,
is important to reduce the number of unsuccessful wells explored.
We found that an optimal value is near $\Delta E_{R} = 3$ Kcal/mol.

It is well known that the local structure of liquid water at equilibrium
consists of a hydrogen bond network formed by oxygen and hydrogen atoms from
neighbouring water molecules.  This structure makes it very hard for an
incoming water molecule to find low energy configurations by forming hydrogen
bonds with pre-existing molecules.  However, the insertion algorithm needs
only to control the thermodynamics by inputting into the system a specified
amount of energy which depends on the ensemble considered.  We performed an MD
simulation of bulk water using a simple spherical water shell to show that it
is possible to insert water molecules on-the-fly while precisely controlling the energy
released to the system. In a previous work \cite{usher} considering Lennard-Jones atoms,
 it was shown that this procedure  ensures thermodynamic consistency after
a relaxation time of the order of the collision time. 
 We set up an equilibrated TIP3P bulk water
system within a sphere of radius $37.5$ {\AA} at 300K and a pressure of 1
atm. The simulations were run  with a 12 {\AA} cutoff radius 
 and without corrections to the  long ranged electrostatic forces \cite{NAMD}. 
 The water molecules in the outer shell of length $d=12.5$ {\AA} play the
role of a reservoir confined in the sphere by a simple constant radial force
field specified by an acceleration ${\bf g}$ acting only within the outer
shell.  The effect of this force is a linear decay of the pressure in the
water shell according to the usual formula for the hydrostatic pressure in an
incompressible fluid $P_{1} = P_{0} - \rho g d $, where $P_{1}$ is the
pressure at the surface of the water sphere and $P_{0} $ is the pressure of
the bulk that we want to maintain.  We impose $P_{1}=0$ by setting
$g=P_{0}/(\rho d)$. 

 In the present set up, the flow rate of molecules to the
inner shell is controlled by the applied pressure force, while the number of
reservoir molecules in the outer shell is fixed at the bulk density.  This
implies that molecules which, due to fluctuations or sudden pressure waves,
move outside the sphere are removed and reinserted using the insertion method
at a random location in the outer shell, with a velocity given by the
Maxwell-Boltzmann distribution at 300K.  We note that the present setting can be
generalised to avoid finite-size effects due to periodic boundary conditions
in a hydrodynamically consistent way\cite{BusH1}.
The total energy of the system can be fixed by setting 
 the amount of energy released upon
insertion  equal to the energy lost  when a molecule moves
out through the open boundary \cite{BusH1}.
On average, the exchanged potential energy per molecule is equal to the mean
energy per molecule: 
by inserting at this energy target we kept the total energy under control (without drift)
with {\it no thermostat}
at all.  In other situations, such as at constant temperature, it is
sufficient to release a moderately greater energy, for example equal to the
excess chemical potential, which can be thermalized dynamically by the
thermostat.

An estimate of the efficiency of this insertion method can be obtained by
 determining the average number of energy evaluations, including failed well searches,
 needed to insert a single water molecule at the specified energy.  
 Each iteration of the insertion algorithm corresponds to one energy evaluation on the solvent 
molecule, which  is a three atom-force calculation for TIP3P water. 
 In particular, it takes an average of $206$ iterations, exploring $34$ wells, 
 to insert at the reference energy of 
 the mean energy per molecule ($-11$ Kcal/mol), and $36$ iterations (only $6$ wells)
 at the energy of the excess chemical
 potential (-5.8 Kcal/mol, calculated using the Bennett method \cite{bennett76}).  
We note that the computational cost required by the insertion method in a typical MD simulation
is quite small. For instance, in the simulation of the open water shell mentioned above,
incoming water molecules were inserted at a target energy of $E_T=-11$Kcal/mol
within a volume of $155.4$nm$^3$ at a rate of $141$ per picosecond. The 
amount of CPU time devoted to insertion was only $3\%$ of the grand total of the simulation.

 \begin{figure}[tb!]
\begin{center}
\centerline{\psfig{figure=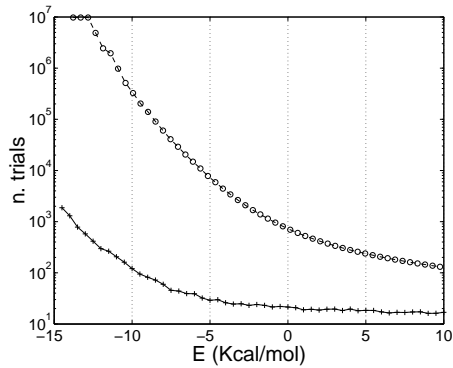,width=6cm} }
\end{center}
\caption{ 
Number of energy evaluations per molecule required to insert a water molecule 
while releasing  an energy less than E (Kcal/mol) to the system. 
The proposed insertion algorithm (crosses) is around three orders of magnitude faster 
than random insertion (circles) at low energies.  
The histogram for  random insertion is computed from $10^7$ trials.
 }
\label{speedup}
\end{figure}

Interestingly, the mean number of iterations to explore a
 well which leads to the correct target energy is only around $12$, independent of
 the target energy.  The method may be improved further by reducing the total
 number of searched wells but it is already optimal in the sense that the
 number of iterations to explore a single well does not depend on the target energy.
 Future applications may require searching many more degrees of freedom,
 e.g. conformational searches, for which it is impractical to fix each maximum
 displacement {\it a priori}. In this case, it would be useful to set up an
 adaptive rule to infer the input parameters from the efficacy of the search
 itself.

It is useful to compare our insertion algorithm with a direct random
insertion.  To this end, the probability distribution $f(E)$ of releasing a
total energy $E$ upon random insertion was estimated by computing a histogram from
$10^7$ random insertion trials.  The number of trials required to obtain an
energy smaller than E is given by the reciprocal of the cumulative distribution
$1/F(E)$ where $F(E)=\int_{-\infty}^{E} f(E')dE'$.  This number is compared
with the number of iterations (energy evaluations) required by the insertion
algorithm in Fig. (\ref{speedup}).  The insertion algorithm is around three
orders of magnitude faster than a random insertion for energies lower than the
chemical potential and so may  
 provide an efficient alternative to biased methods, such
as cavity-biased sampling \cite{jed00,poh96}, to reconstruct the probability
distribution $f(E)$. Indeed, the
present algorithm enables one to identify the important low energy regions  
very accurately where an un-biased sampling can be performed. 
This  appealing approach enables   fast computation of the
chemical potential from the probability distribution $f(E)$
at low energies\cite{bus05}.


In summary, we have reported a new method for the insertion of polar molecules
 in dense fluids by a generalisation of the {\sc usher} protocol \cite{usher}.
The energy minimisation is applied concurrently
to all degrees of freedom (translational and rotational for water) and is
independent of the specific potential used. Indeed, the method is even more
general. It may be applied to other problems related to conformational
searches and minima of  potential energy surfaces with many more degrees
freedom. Given its importance for computational biology, we  focused on water and 
 demonstrated that it is possible to efficiently insert water molecules in
 aqueous environments while controlling the thermodynamic state.  
This task is commonly considered to be very time consuming, 
but we are able to  achieve it at  negligible computational cost thanks to 
a very efficient configurational search algorithm. 
The present algorithm is an essential tool for performing hybrid MD-continuum
simulations \cite{BusH1,SandraBus} of biological interest. Indeed, it 
represents an important step towards a general method for performing MD simulations of
open systems, for which a dynamic calculation of the chemical potential 
 \cite{lu03,bus05} could 
be used to control the insertion rate so as to maintain constant the solvent chemical
potential.

This research was supported by the EPSRC Integrative Biology project GR/S72023 and by the EPSRC RealityGrid project GR/67699.

\bibliographystyle{apsrev}
\bibliography{water.bib}

\begin{thebibliography}{29}
\expandafter\ifx\csname natexlab\endcsname\relax\def\natexlab#1{#1}\fi
\expandafter\ifx\csname bibnamefont\endcsname\relax
  \def\bibnamefont#1{#1}\fi
\expandafter\ifx\csname bibfnamefont\endcsname\relax
  \def\bibfnamefont#1{#1}\fi
\expandafter\ifx\csname citenamefont\endcsname\relax
  \def\citenamefont#1{#1}\fi
\expandafter\ifx\csname url\endcsname\relax
  \def\url#1{\texttt{#1}}\fi
\expandafter\ifx\csname urlprefix\endcsname\relax\def\urlprefix{URL }\fi
\providecommand{\bibinfo}[2]{#2}
\providecommand{\eprint}[2][]{\url{#2}}

\bibitem[{\citenamefont{Allen and Tildesley}(1987)}]{allen}
\bibinfo{author}{\bibfnamefont{M.}~\bibnamefont{Allen}} \bibnamefont{and}
  \bibinfo{author}{\bibfnamefont{D.}~\bibnamefont{Tildesley}},
  \emph{\bibinfo{title}{Computer Simulations of Liquids}}
  (\bibinfo{publisher}{Oxford University Press}, \bibinfo{year}{1987}).

\bibitem[{\citenamefont{Adams}(1975)}]{adams75}
\bibinfo{author}{\bibfnamefont{D.}~\bibnamefont{Adams}}, \bibinfo{journal}{Mol.
  Phys.} \textbf{\bibinfo{volume}{29}}, \bibinfo{pages}{307}
  (\bibinfo{year}{1975}).

\bibitem[{\citenamefont{Mezei}(1987)}]{Mezei87}
\bibinfo{author}{\bibfnamefont{M.}~\bibnamefont{Mezei}}, \bibinfo{journal}{Mol.
  Phys.} \textbf{\bibinfo{volume}{61}}, \bibinfo{pages}{565}
  (\bibinfo{year}{1987}).

\bibitem[{\citenamefont{Jie~Ji and Pettitt}(1992)}]{Pet92}
\bibinfo{author}{\bibfnamefont{T.~C.} \bibnamefont{Jie~Ji}} \bibnamefont{and}
  \bibinfo{author}{\bibfnamefont{B.~M.} \bibnamefont{Pettitt}},
  \bibinfo{journal}{J. Chem. Phys.} \textbf{\bibinfo{volume}{96}},
  \bibinfo{pages}{1333} (\bibinfo{year}{1992}).

\bibitem[{\citenamefont{G.C.~Lynch}(2000)}]{Pet-bov}
\bibinfo{author}{\bibfnamefont{B.}~\bibnamefont{G.C.~Lynch}},
  \bibinfo{journal}{Chemical Physics} \textbf{\bibinfo{volume}{258}},
  \bibinfo{pages}{405} (\bibinfo{year}{2000}).

\bibitem[{\citenamefont{Delgado-Buscalioni and
  Coveney}(2003{\natexlab{a}})}]{BusH1}
\bibinfo{author}{\bibfnamefont{R.}~\bibnamefont{Delgado-Buscalioni}}
  \bibnamefont{and} \bibinfo{author}{\bibfnamefont{P.~V.}
  \bibnamefont{Coveney}}, \bibinfo{journal}{Phys. Rev. E}
  \textbf{\bibinfo{volume}{67}}, \bibinfo{pages}{046704}
  (\bibinfo{year}{2003}{\natexlab{a}}).

\bibitem[{\citenamefont{Flekk{\o}y et~al.}(2000)\citenamefont{Flekk{\o}y,
  Wagner, and Feder}}]{Flek00}
\bibinfo{author}{\bibfnamefont{E.~G.} \bibnamefont{Flekk{\o}y}},
  \bibinfo{author}{\bibfnamefont{G.}~\bibnamefont{Wagner}}, \bibnamefont{and}
  \bibinfo{author}{\bibfnamefont{J.}~\bibnamefont{Feder}},
  \bibinfo{journal}{Europhys. Lett.} \textbf{\bibinfo{volume}{52}},
  \bibinfo{pages}{271} (\bibinfo{year}{2000}).

\bibitem[{\citenamefont{Barsky et~al.}(2004)\citenamefont{Barsky,
  Delgado-Buscalioni, and Coveney}}]{SandraBus}
\bibinfo{author}{\bibfnamefont{S.}~\bibnamefont{Barsky}},
  \bibinfo{author}{\bibfnamefont{R.}~\bibnamefont{Delgado-Buscalioni}},
  \bibnamefont{and} \bibinfo{author}{\bibfnamefont{P.~V.}
  \bibnamefont{Coveney}}, \bibinfo{journal}{J. Chem. Phys.}
  \textbf{\bibinfo{volume}{121}}, \bibinfo{pages}{2403} (\bibinfo{year}{2004}).

\bibitem[{\citenamefont{King and Warshel}(1989)}]{SCAAS}
\bibinfo{author}{\bibfnamefont{G.}~\bibnamefont{King}} \bibnamefont{and}
  \bibinfo{author}{\bibfnamefont{A.}~\bibnamefont{Warshel}},
  \bibinfo{journal}{J. Chem. Phys.} \textbf{\bibinfo{volume}{91}},
  \bibinfo{pages}{3647} (\bibinfo{year}{1989}).

\bibitem[{\citenamefont{Goodfellow et~al.}(1996)\citenamefont{Goodfellow,
  Knaggs, Williams, and Thornton}}]{Goodfellow1}
\bibinfo{author}{\bibfnamefont{J.~M.} \bibnamefont{Goodfellow}},
  \bibinfo{author}{\bibfnamefont{M.}~\bibnamefont{Knaggs}},
  \bibinfo{author}{\bibfnamefont{M.~A.} \bibnamefont{Williams}},
  \bibnamefont{and} \bibinfo{author}{\bibfnamefont{J.~M.}
  \bibnamefont{Thornton}}, \bibinfo{journal}{Faraday Discussions}
  \textbf{\bibinfo{volume}{103}}, \bibinfo{pages}{339} (\bibinfo{year}{1996}).

\bibitem[{\citenamefont{M.A.Williams et~al.}(1997)\citenamefont{M.A.Williams,
  J.M.Thornton, and J.M.Goodfellow}}]{Goodfellow2}
\bibinfo{author}{\bibnamefont{M.A.Williams}},
  \bibinfo{author}{\bibnamefont{J.M.Thornton}}, \bibnamefont{and}
  \bibinfo{author}{\bibnamefont{J.M.Goodfellow}}, \bibinfo{journal}{Protein
  Engineering} \textbf{\bibinfo{volume}{10}}, \bibinfo{pages}{895}
  (\bibinfo{year}{1997}).

\bibitem[{\citenamefont{Zhang and Hermans}(1996)}]{Hermans96}
\bibinfo{author}{\bibfnamefont{L.}~\bibnamefont{Zhang}} \bibnamefont{and}
  \bibinfo{author}{\bibfnamefont{J.}~\bibnamefont{Hermans}},
  \bibinfo{journal}{Proteins:Struct. Func. Genet.}
  \textbf{\bibinfo{volume}{24}}, \bibinfo{pages}{433} (\bibinfo{year}{1996}).

\bibitem[{\citenamefont{Hofacker and Schulten}(1998)}]{Ivo98}
\bibinfo{author}{\bibfnamefont{I.}~\bibnamefont{Hofacker}} \bibnamefont{and}
  \bibinfo{author}{\bibfnamefont{K.}~\bibnamefont{Schulten}},
  \bibinfo{journal}{Proteins:Struct. Func. Genet.}
  \textbf{\bibinfo{volume}{30}}, \bibinfo{pages}{100} (\bibinfo{year}{1998}).

\bibitem[{\citenamefont{Jensen et~al.}(2003)\citenamefont{Jensen, Tajkorshid,
  and Schulten}}]{AQP-03}
\bibinfo{author}{\bibfnamefont{M.}~\bibnamefont{Jensen}},
  \bibinfo{author}{\bibfnamefont{E.}~\bibnamefont{Tajkorshid}},
  \bibnamefont{and} \bibinfo{author}{\bibfnamefont{K.}~\bibnamefont{Schulten}},
  \bibinfo{journal}{Biophysical Journal} \textbf{\bibinfo{volume}{85}},
  \bibinfo{pages}{2884} (\bibinfo{year}{2003}).

\bibitem[{\citenamefont{Cai et~al.}(2003)\citenamefont{Cai, Stevens, Budor, and
  Zuiderweg}}]{cai03biochemistry}
\bibinfo{author}{\bibfnamefont{S.}~\bibnamefont{Cai}},
  \bibinfo{author}{\bibfnamefont{S.}~\bibnamefont{Stevens}},
  \bibinfo{author}{\bibfnamefont{A.}~\bibnamefont{Budor}}, \bibnamefont{and}
  \bibinfo{author}{\bibfnamefont{E.}~\bibnamefont{Zuiderweg}},
  \bibinfo{journal}{Biochemistry} \textbf{\bibinfo{volume}{42}},
  \bibinfo{pages}{9} (\bibinfo{year}{2003}).

\bibitem[{\citenamefont{Bennett}(1976)}]{bennett76}
\bibinfo{author}{\bibfnamefont{C.~H.} \bibnamefont{Bennett}},
  \bibinfo{journal}{J. Comput. Phys.} \textbf{\bibinfo{volume}{22}},
  \bibinfo{pages}{245} (\bibinfo{year}{1976}).

\bibitem[{\citenamefont{K.S.Shing and K.E.Gubbins}(1982)}]{shing82}
\bibinfo{author}{\bibnamefont{K.S.Shing}} \bibnamefont{and}
  \bibinfo{author}{\bibnamefont{K.E.Gubbins}}, \bibinfo{journal}{Mol. Phys.}
  \textbf{\bibinfo{volume}{46}}, \bibinfo{pages}{1109} (\bibinfo{year}{1982}).

\bibitem[{\citenamefont{Lu et~al.}(2003)\citenamefont{Lu, Singh, and
  Kofke}}]{lu03}
\bibinfo{author}{\bibfnamefont{N.}~\bibnamefont{Lu}},
  \bibinfo{author}{\bibfnamefont{J.~K.} \bibnamefont{Singh}}, \bibnamefont{and}
  \bibinfo{author}{\bibfnamefont{D.~A.} \bibnamefont{Kofke}},
  \bibinfo{journal}{J. Chem. Phys.} \textbf{\bibinfo{volume}{118}},
  \bibinfo{pages}{2977} (\bibinfo{year}{2003}).

\bibitem[{\citenamefont{Guillot et~al.}(1991)\citenamefont{Guillot, Guissani,
  and Bratos}}]{guillot91}
\bibinfo{author}{\bibfnamefont{B.}~\bibnamefont{Guillot}},
  \bibinfo{author}{\bibfnamefont{Y.}~\bibnamefont{Guissani}}, \bibnamefont{and}
  \bibinfo{author}{\bibfnamefont{S.}~\bibnamefont{Bratos}},
  \bibinfo{journal}{J. Chem. Phys.} \textbf{\bibinfo{volume}{95}},
  \bibinfo{pages}{3643} (\bibinfo{year}{1991}).

\bibitem[{\citenamefont{Jedlovszky and Mezei}(2000)}]{jed00}
\bibinfo{author}{\bibfnamefont{P.}~\bibnamefont{Jedlovszky}} \bibnamefont{and}
  \bibinfo{author}{\bibfnamefont{M.}~\bibnamefont{Mezei}}, \bibinfo{journal}{J.
  Am. Chem. Soc.} \textbf{\bibinfo{volume}{122}}, \bibinfo{pages}{5125}
  (\bibinfo{year}{2000}).

\bibitem[{\citenamefont{Pohorille and Wilson}(1996)}]{poh96}
\bibinfo{author}{\bibfnamefont{A.}~\bibnamefont{Pohorille}} \bibnamefont{and}
  \bibinfo{author}{\bibfnamefont{M.~A.} \bibnamefont{Wilson}},
  \bibinfo{journal}{J. Chem. Phys.} \textbf{\bibinfo{volume}{104}},
  \bibinfo{pages}{3760} (\bibinfo{year}{1996}).

\bibitem[{\citenamefont{G.L.Deitrick et~al.}(1989)\citenamefont{G.L.Deitrick,
  Scriven, and Davis}}]{dei89}
\bibinfo{author}{\bibnamefont{G.L.Deitrick}},
  \bibinfo{author}{\bibfnamefont{L.}~\bibnamefont{Scriven}}, \bibnamefont{and}
  \bibinfo{author}{\bibfnamefont{H.}~\bibnamefont{Davis}}, \bibinfo{journal}{J.
  Chem. Phys.} \textbf{\bibinfo{volume}{90}}, \bibinfo{pages}{2370}
  (\bibinfo{year}{1989}).

\bibitem[{\citenamefont{Delgado-Buscalioni and
  Coveney}(2003{\natexlab{b}})}]{usher}
\bibinfo{author}{\bibfnamefont{R.}~\bibnamefont{Delgado-Buscalioni}}
  \bibnamefont{and} \bibinfo{author}{\bibfnamefont{P.~V.}
  \bibnamefont{Coveney}}, \bibinfo{journal}{J. Chem. Phys.}
  \textbf{\bibinfo{volume}{119}}, \bibinfo{pages}{978}
  (\bibinfo{year}{2003}{\natexlab{b}}).

\bibitem[{\citenamefont{Levitt and Warshel}(1975)}]{levitt75}
\bibinfo{author}{\bibfnamefont{M.}~\bibnamefont{Levitt}} \bibnamefont{and}
  \bibinfo{author}{\bibfnamefont{A.}~\bibnamefont{Warshel}},
  \bibinfo{journal}{Nature} \textbf{\bibinfo{volume}{253}},
  \bibinfo{pages}{694} (\bibinfo{year}{1975}).

\bibitem[{\citenamefont{Li and Scheraga}(1987)}]{li87}
\bibinfo{author}{\bibfnamefont{Z.}~\bibnamefont{Li}} \bibnamefont{and}
  \bibinfo{author}{\bibfnamefont{H.~A.} \bibnamefont{Scheraga}},
  \bibinfo{journal}{Proc. Natl. Acad. Sci. USA} \textbf{\bibinfo{volume}{84}},
  \bibinfo{pages}{6611} (\bibinfo{year}{1987}).

\bibitem[{\citenamefont{Saunders}(1987)}]{saunders87}
\bibinfo{author}{\bibfnamefont{M.}~\bibnamefont{Saunders}},
  \bibinfo{journal}{J. Am. Chem. Soc.} \textbf{\bibinfo{volume}{109}},
  \bibinfo{pages}{3150} (\bibinfo{year}{1987}).

\bibitem[{\citenamefont{Kal\'e et~al.}(1999)\citenamefont{Kal\'e, Skeel,
  Bhandarkar, Brunner, Gursoy, Krawetz, Phillips, Shinozaki, Varadarajan, and
  Schulten}}]{NAMD}
\bibinfo{author}{\bibfnamefont{L.}~\bibnamefont{Kal\'e}},
  \bibinfo{author}{\bibfnamefont{R.}~\bibnamefont{Skeel}},
  \bibinfo{author}{\bibfnamefont{M.}~\bibnamefont{Bhandarkar}},
  \bibinfo{author}{\bibfnamefont{R.}~\bibnamefont{Brunner}},
  \bibinfo{author}{\bibfnamefont{A.}~\bibnamefont{Gursoy}},
  \bibinfo{author}{\bibfnamefont{N.}~\bibnamefont{Krawetz}},
  \bibinfo{author}{\bibfnamefont{J.}~\bibnamefont{Phillips}},
  \bibinfo{author}{\bibfnamefont{A.}~\bibnamefont{Shinozaki}},
  \bibinfo{author}{\bibfnamefont{K.}~\bibnamefont{Varadarajan}},
  \bibnamefont{and} \bibinfo{author}{\bibfnamefont{K.}~\bibnamefont{Schulten}},
  \bibinfo{journal}{J. Comp. Phys.} \textbf{\bibinfo{volume}{151}},
  \bibinfo{pages}{283} (\bibinfo{year}{1999}).

\bibitem[{\citenamefont{Jorgensen et~al.}(1983)\citenamefont{Jorgensen,
  Chandrasekhar, and Madura}}]{jorgensen83}
\bibinfo{author}{\bibfnamefont{W.~L.} \bibnamefont{Jorgensen}},
  \bibinfo{author}{\bibfnamefont{J.}~\bibnamefont{Chandrasekhar}},
  \bibnamefont{and} \bibinfo{author}{\bibfnamefont{D.}~\bibnamefont{Madura}},
  \bibinfo{journal}{J. Chem. Phys.} \textbf{\bibinfo{volume}{79}},
  \bibinfo{pages}{926} (\bibinfo{year}{1983}).

\bibitem[{\citenamefont{Delgado-Buscalioni
  et~al.}(2004)\citenamefont{Delgado-Buscalioni, {De Fabritiis}, and
  Coveney}}]{bus05}
\bibinfo{author}{\bibfnamefont{R.}~\bibnamefont{Delgado-Buscalioni}},
  \bibinfo{author}{\bibfnamefont{G.}~\bibnamefont{{De Fabritiis}}},
  \bibnamefont{and} \bibinfo{author}{\bibfnamefont{P.}~\bibnamefont{Coveney}},
  \bibinfo{journal}{Fast calculation of the chemical potential using
  energy-biased sampling, preprint}  (\bibinfo{year}{2004}).

\end{thebibliography}


%
%
%
%
%

\end{document}